\documentclass{cs19proc}

\editors{G.~A. Feiden}
\publisher{Zenodo}
\conference{The 19th Cambridge Workshop on Cool Stars, Stellar Systems, and the Sun}
\conferencedate{2016}

\title{The coronal evolution of pre-main-sequence stars}
\author{Scott G. Gregory,$^{1}$ 
            Fred C. Adams,$^{2,3}$
            Claire L. Davies$^{4,1}$}

\affiliation{$^{1}$School of Physics and Astronomy, University of St Andrews, St Andrews, KY16 9SS, UK \\
	$^{2}$Astronomy Department, University of Michigan, Ann Arbor, MI 48109, USA  \\ 
	$^{3}$Physics Department, University of Michigan, Ann Arbor, MI 48109, USA \\
	$^{4}$School of Physics, University of Exeter, Exeter EX4 4QL, UK }

\shorttitle{The evolution of PMS stars}
\shortauthors{S. G. Gregory, F. C. Adams \& C. L. Davies}

\abs{The bulk of X-ray emission from pre-main-sequence (PMS) stars is coronal in origin. We demonstrate herein that stars on Henyey tracks in the Hertzsprung-Russell diagram have lower $\log(L_X/L_\ast)$, on average, than stars on Hayashi tracks. This effect is driven by the decay of $L_X$ once stars develop radiative cores.  $L_X$ decays faster with age for intermediate mass PMS stars, the progenitors of main sequence A-type stars, compared to those of lower mass.  As almost all main sequence A-type stars show no detectable X-ray emission, we may already be observing the loss of their coronae during their PMS evolution.  Although there is no direct link between the size or mass of the radiative core and $L_X$, the longer stars have spent with partially convective interiors, the weaker their X-ray emission becomes.  This conference paper is a synopsis of Gregory, Adams \& Davies (2016).}

\begin{document}

\maketitle


\section{Introduction}\label{intro}
Copious coronal X-ray emission is a defining characteristic of pre-main-sequence (PMS) stars.  With X-ray luminosities of $\log L_X \approx$ 28--32 (with $L_X$ in units of ${\rm erg\,s^{-1}}$) and an average coronal temperature of $\approx$30$\,{\rm MK}$ \citep{pre05}, they are far more X-ray luminous and hotter than the contemporary Sun ($\log L_X \approx$26.4--27.7 at solar minimum and maximum, with a coronal temperature of $\approx$2$\,{\rm MK}$; \citealt{per00}).      

PMS star X-ray luminosity increases with stellar mass and decreases with age (e.g. \citealt{pre05,pre05age}), albeit with large scatter in $L_X$ at any given mass or age. However, the stellar rotation rate is less important for PMS stellar X-ray emission, at least for members of the youngest star forming regions.  PMS stars typically all show saturated levels of X-ray emission unlike low-mass stars in main sequence clusters which follow the rotation-activity relation (e.g. \citealt{wri11}). Recently, \citet{arg16} demonstrated that, within a sample of stars over a certain mass range in the intermediate age PMS cluster h~Per, the slow rotators have begun to show evidence for unsaturated X-ray emission, and the fast rotators supersaturation.

In this work we investigate how the stellar internal structure influences the coronal X-ray emission.  It is known that the large-scale magnetic topology of PMS stars is linked to the evolution of the stellar internal structure \citep{gre12,gre14,fol16}.  More evolved PMS stars (those with large radiative cores) are found to have more complex, multipolar, and non-axisymmetric magnetic fields compared to less evolved stars (at least for those more massive than $\sim$0.5$\,{\rm M}_\odot$, as little is known about the field topology of lower mass PMS stars; \citealt{gre12}).  An increase in the large-scale magnetic field complexity likely corresponds to a reduction in the available X-ray emitting volume, with stellar coronae becoming more compact. We therefore expect that there may be differences in the X-ray luminosities of fully and partially convective PMS stars.  That is indeed what we find and discuss in this paper.

In \S\ref{evolution} we briefly discuss the physics of radiative core development during the PMS contraction.  In \S\ref{decay} we discuss our sample of PMS stars collated from the literature.  We derive stellar masses, ages, and internal structure information from the models of \citet{sie00}. We then continue by comparing the X-ray luminosities, and the temporal evolution of $L_X$, for fully and partially convective PMS stars, and for stars on Hayashi and Henyey tracks in the Hertzsprung-Russell (HR) diagram. Although in this paper we only consider the \citet{sie00} models, in \citet{gre16} we considered three different PMS evolutionary models. All of the general results and conclusions discussed here remain the same, regardless of model choice. In \S\ref{Atype} we argue that the lack of X-ray detections from main sequence A-type stars is a result of stars losing their coronae as the depth of their convective zone reduces during their PMS evolution. We conclude in \S\ref{conclusions}.


\section{The evolution of PMS stars across the HR diagram}\label{evolution}
When low-mass PMS stars first become optically visible they are located in the upper-right of the $\log(L_\ast/{\rm L}_\odot)$ vs $\log T_{\rm eff}$ HR diagram.  From there, stars of different mass follow different paths across the HR diagram as they contract under gravity.  Initially, they host fully convective interiors as they evolve along Hayashi tracks, where their surface luminosity, $L_\ast$, decreases with increasing age.  The internal luminosity of the star increases monotonically from the centre of the star to the surface value whilst the interior is fully convective.  

As contraction proceeds, the temperature increases in the core and the opacity drops, eventually inhibiting convection in the central regions if the star is massive enough (stars of mass $\lesssim$0.35$\,{\rm M}_\odot$ remain fully convective for their entire evolution). The now radiative core grows outwards at the expense of the convective zone depth. Mass shells interior to the core-envelope boundary lose heat, while those exterior to the boundary gain heat.  The base of the convective zone is heated and the interior luminosity of the star rises from the centre to a maximum before dropping towards the surface.  This internal luminosity maximum radiatively diffuses towards the stellar surface, and eventually the surface luminosity, $L_\ast$, begins to increase with increasing age as stars evolve onto Henyey tracks.  Note that stars of mass $\lesssim$0.65$\,{\rm M}_\odot$ reach the ZAMS while still on Hayashi tracks. By the time stars evolve onto Henyey tracks, the bulk of the stellar mass is contained within the radiative core ($\sim$60-70\%; \citealt{sie00}), which occupies a smaller proportion of the star by radius. Thus, stars are more centrally condensed by the time their surface luminosity has begun to increase.  

\begin{figure}
	\centering
	\includegraphics[width=0.8\linewidth]{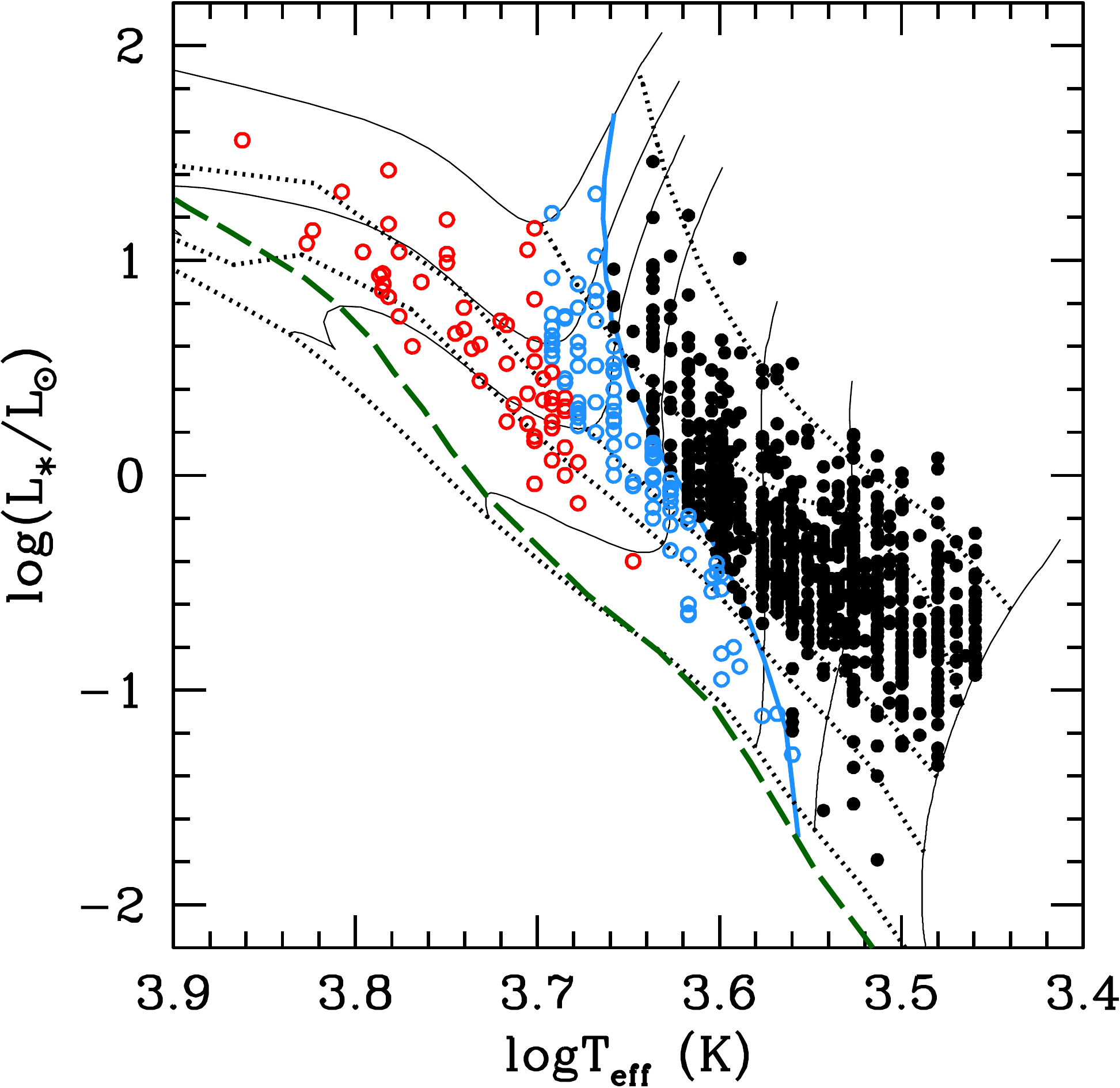}
	\caption{A Hertzsprung-Russell diagram showing the PMS stars considered in this work.  The mass tracks (solid black lines for 0.1, 0.3, 0.5, 1, 1.5, 2, and 3$\,{\rm M}_\odot$) and isochrones (dotted lines for 0.1, 1, 5, 10, and 60$\,{\rm Myr}$) are from \citet{sie00}, $Z=0.02$ with convective overshooting. The dashed green line is the ZAMS. The solid blue line divides stars with fully convective interiors (black points) from those that have developed radiative cores. The blue and red points denote partially convective stars with radiative cores on Hayashi tracks (stars with $L_\ast$ decreasing with increasing age) and Henyey tracks (stars with $L_\ast$ increasing with increasing age), respectively.}
	\label{HRdiagram}
\end{figure}

In this work we compare the coronal X-ray emission of fully and partially convective PMS stars, and of Hayashi track and Henyey track PMS stars.  Stars on Hayashi tracks can be either fully or partially convective, while those on Henyey tracks have mostly radiative interiors.  The delay between radiative core development and a star evolving onto its Henyey track is significant.  As one example, using the models of \citet{sie00}, a solar mass star becomes partially convective after $\sim$2.5$\,{\rm Myr}$ and evolves onto its Henyey track at $\sim$15$\,{\rm Myr}$.  Therefore, a solar mass star spends longer with a partially convective interior than it does with a fully convective interior while on its Hayashi track. A solar mass star spends $\sim$40\% of its entire PMS lifetime of $\sim$30$\,{\rm Myr}$ with a radiative core on its Hayashi track.

In the following subsection we compare the coronal X-ray emission properties of fully and partially convective PMS stars, and of those on Hayashi tracks to those on Henyey tracks in the HR diagram.


\section{The decay of PMS star X-ray emission}\label{decay}
To compare the X-ray properties of fully and partially convective PMS stars we considered five of the best studied star forming regions: the Orion Nebula Cluster (ONC), NGC~6530, NGC~2264, IC~348, and NGC~2362. We obtained X-ray luminosities ($L_X$), spectral types, observed photometry, binary star status, and cluster distance estimates from the literature. We dereddened the photometry using intrinsic colours appropriate for the spectral type, from the PMS calibrated scale of \cite{pec13}.  We then calculated bolometric luminosities by applying a spectral type-dependent bolometric correction and with an assumed distance modulus appropriate for each cluster.  We also assigned effective temperatures from the same scale of \citet{pec13}.  Extensive details can be found in \citet{gre16}. Figure \ref{HRdiagram} is a HR diagram showing the stars in our sample, with mass tracks and isochrones from the models of \citet{sie00}, from which we obtained stellar masses, ages, and internal structure information.

Known or suspected spectroscopic / close binaries were removed from our sample in case of confusion in the optical or X-ray data.  With the exception of IC~348, we used the X-ray luminosities from the MYStIX (Massive Young Star-Forming Complex Study in Infrared and X-ray) project \citep{fei13}, in particular those listed in the MPCM (MYStIX Probable Complex Members) catalog of \citet{bro13}. For IC~348, we took X-ray fluxes from \citet{ste12}.  Our final sample consisted of 984 stars, of which 34 are $L_X$ upper limits (all from IC~348).

\begin{figure}
	\centering
	\includegraphics[width=0.78\linewidth]{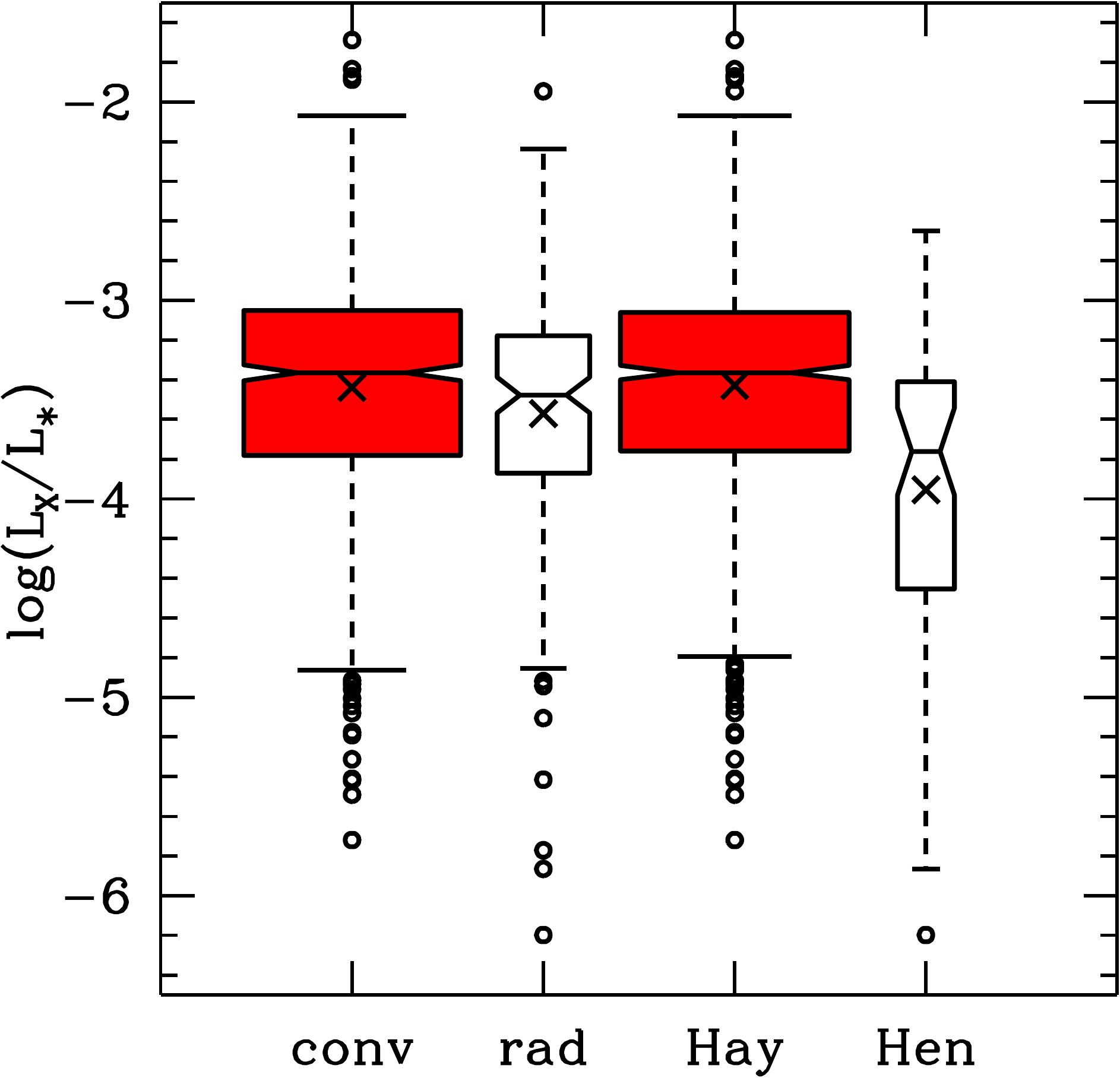}
	\caption{Notched and variable width (scaled to the square root of the sample size in each case) box plots with outliers for the distribution of $\log(L_X/L_\ast)$ for (left-to-right) fully convective (conv), partially convective (rad), Hayashi track (Hay), and Henyey track (Hen) PMS stars. The cross is the mean value in each case. A difference in medians is significant when the notches of the box plots being compared do not overlap (e.g. \citealt{mcg78}). Partially convective PMS stars have, on average, lower $\log(L_X/L_\ast)$ compared to fully convective stars.  The deficit is greater when comparing Hayashi to Henyey track PMS stars.  The difference is caused by the decay of $L_X$ when stars develop substantial radiative cores, see text and Figure \ref{logLX_logLstar}. Figure modified from \citet{gre16}.}
	\label{hist}
\end{figure}

Several previous studies have reported that PMS stars on radiative tracks in the HR diagram have lower average values of $\log(L_X/L_\ast)$ compared to those on convective tracks (e.g. \citealt{fei03,fla03,reb06,cur09,may10}). For stars of mass 1-2$\,{\rm M}_\odot$, \citet{reb06} report a factor of about 10 reduction in $\log(L_X/L_\ast)$ for partially convective stars compared to fully convective stars.  About 28\% of the stars considered by \citet{reb06} are $L_X$ upper limits.  The re-analysis of archival data during the MYStIX project, using the modern methods of \citet{get10} which allows $L_X$ values to be derived for faint sources that have too few counts for traditional X-ray spectral fitting methods, has eliminated almost all $L_X$ upper limits. This, combined with enhanced spectroscopic surveys of the star-forming regions (e.g. \citealt{hil13}), and new empirical colours / temperature scales calibrated for PMS stars \citep{pec13,her14}, warrants our re-examination of differences in the X-rays properties of fully and partially convective PMS stars.     

In Figure \ref{hist} we compare the distributions of the logarithmic fractional X-ray luminosities, $\log(L_X/L_\ast)$, for fully and partially convective PMS stars, and for stars on Hayashi and Henyey tracks in the HR diagram. There is a large scatter in $\log(L_X/L_\ast)$ for stars of all HR diagram locations / internal structures. However, partially convective PMS stars (those which have developed radiative cores), have $\langle\log(L_X/L_\ast)\rangle$ 0.13 dex lower than what we find for fully convective PMS stars. The deficit is larger when comparing Hayashi track stars to Henyey track stars with the latter having $\langle\log(L_X/L_\ast)\rangle$ 0.52 dex less than the former.  About 50\% of our entire sample are stars from the ONC, one of the youngest star forming regions.  Within the ONC sample, about 80\% of the partially convective stars are still on Hayashi tracks.  If we neglect the ONC stars, and consider the other four star-forming regions only, the $\langle\log(L_X/L_\ast)\rangle$ deficit is larger, with a 0.38 (0.64) dex reduction when comparing fully convective to partially convective (Hayashi track to Henyey track) stars.       

\begin{figure*}
	\centering
	\includegraphics[width=0.33\linewidth]{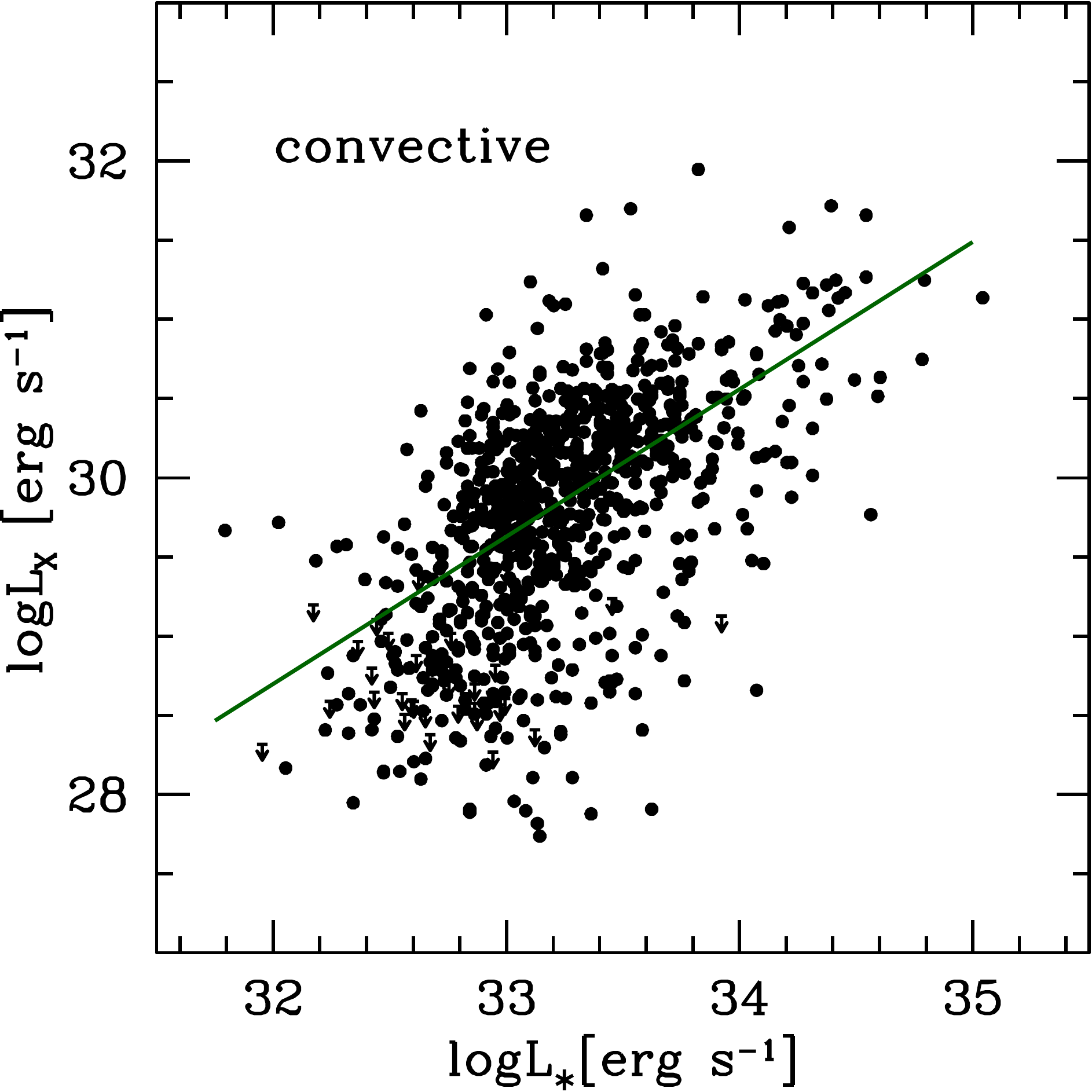}
	\includegraphics[width=0.33\linewidth]{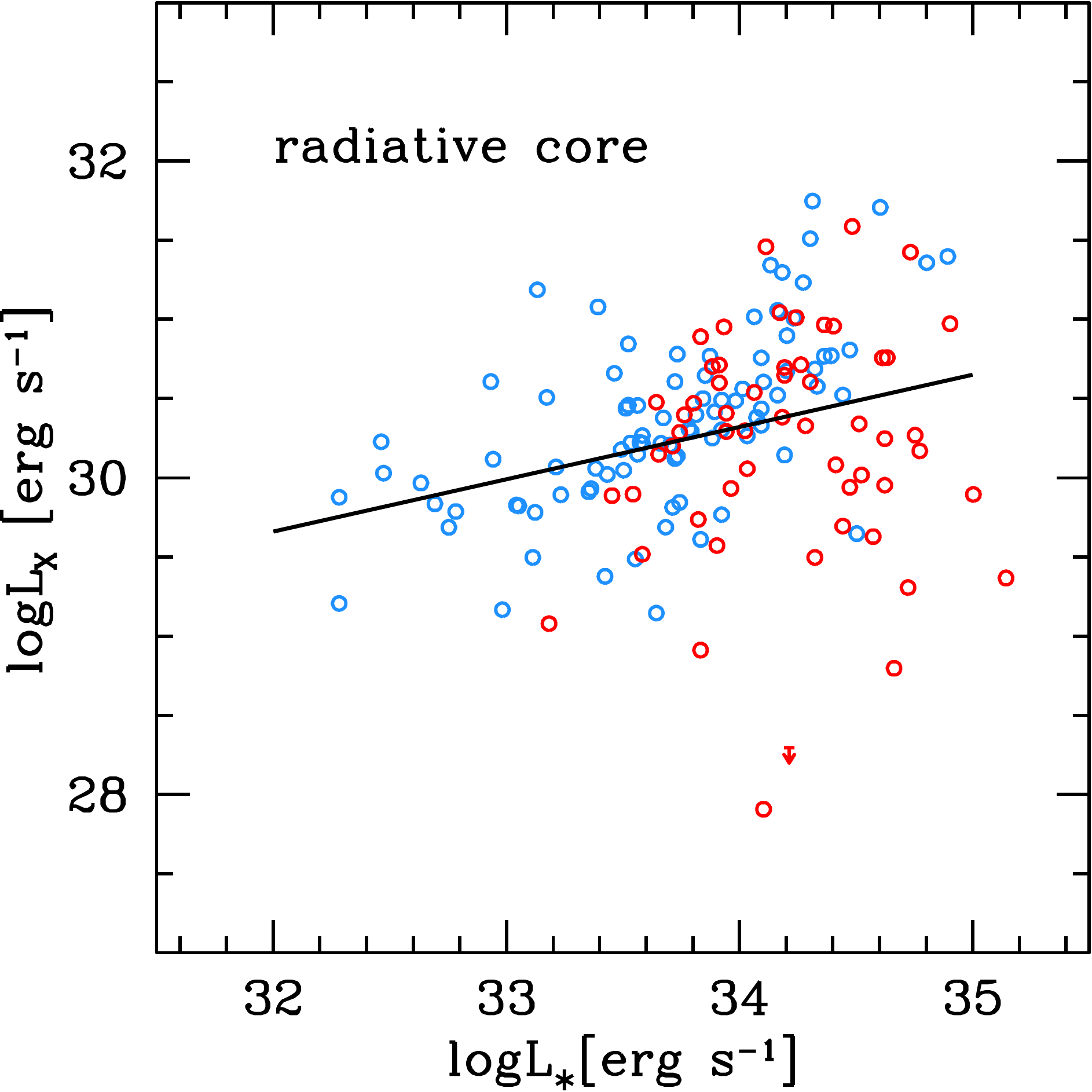}\\
	\includegraphics[width=0.33\linewidth]{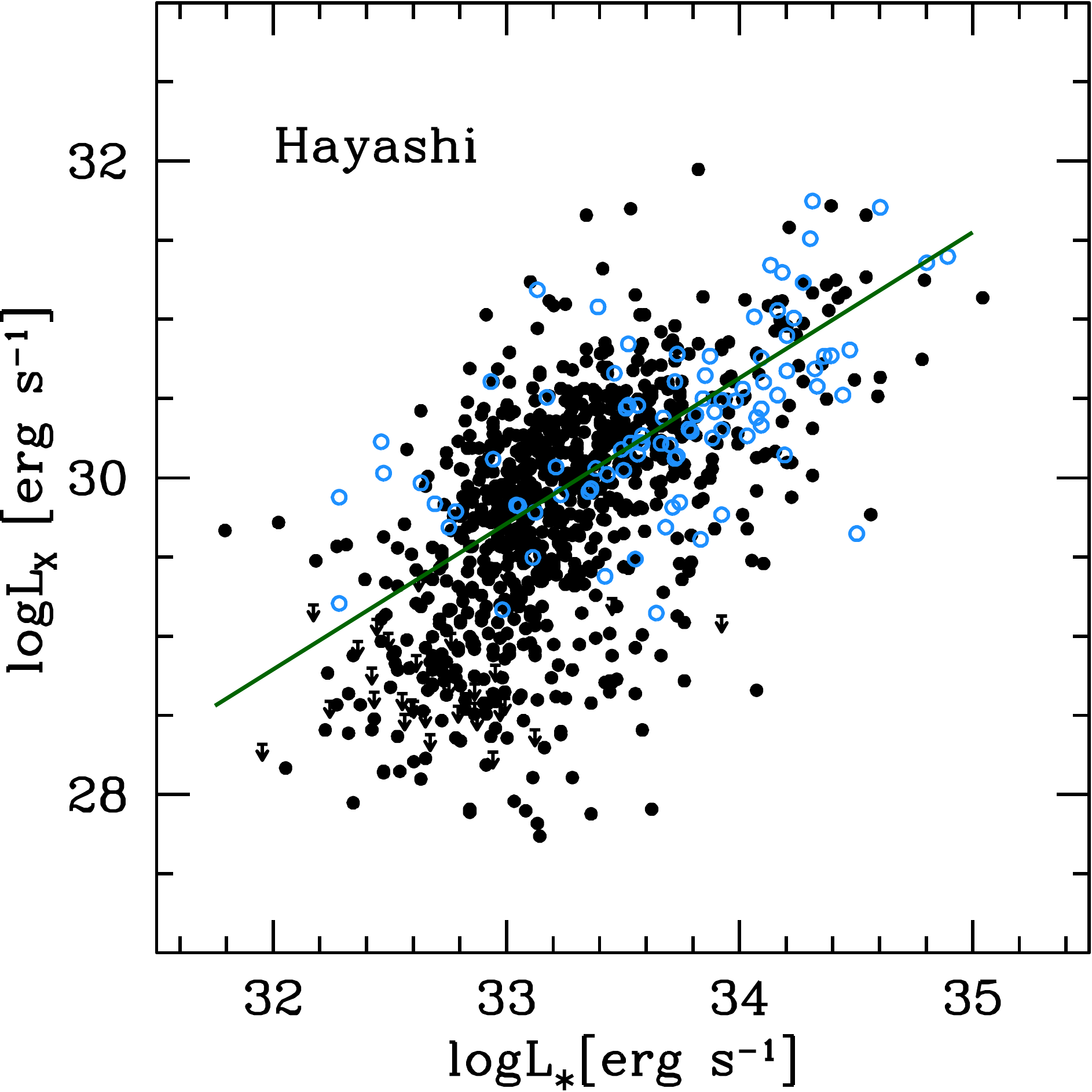}
	\includegraphics[width=0.33\linewidth]{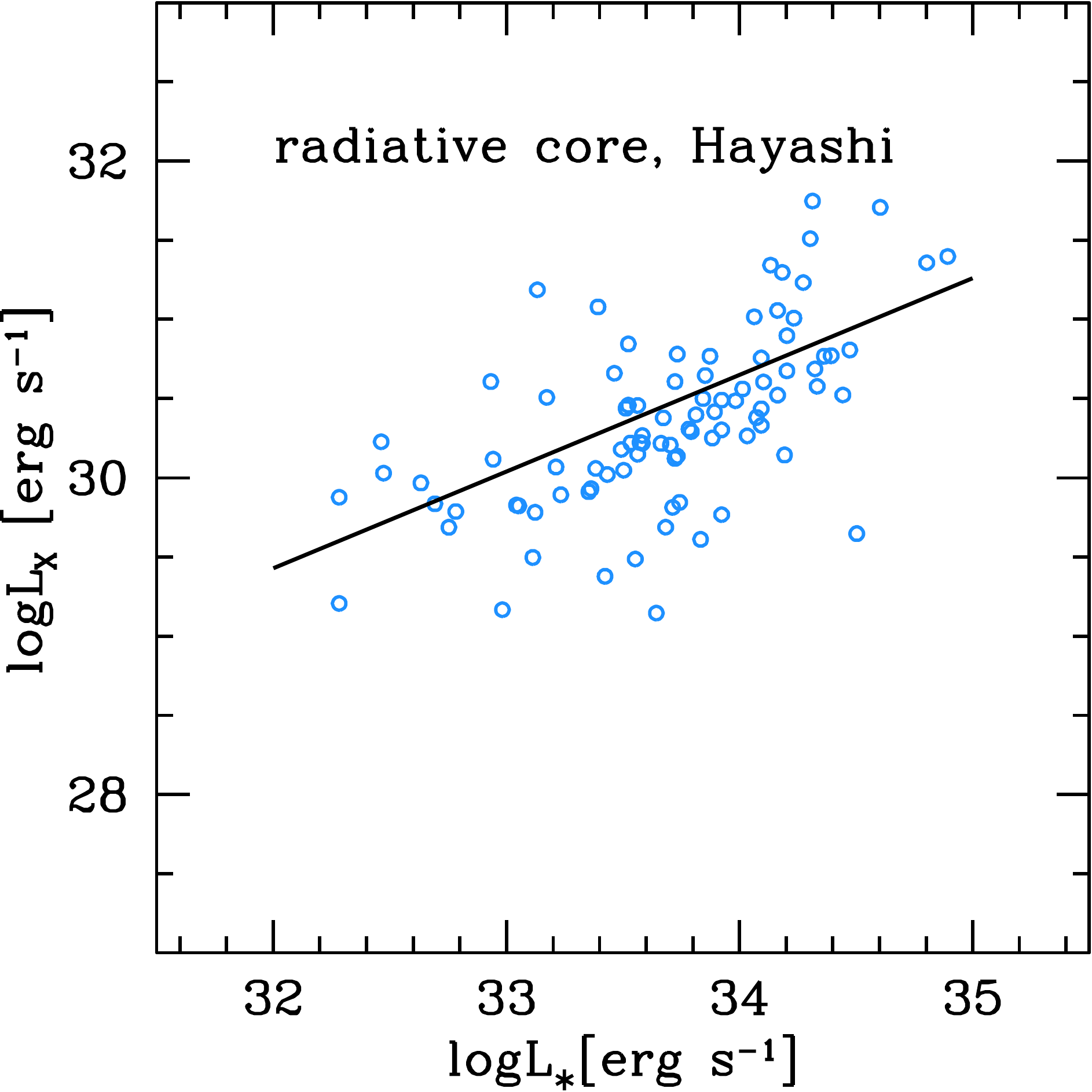}
	\includegraphics[width=0.33\linewidth]{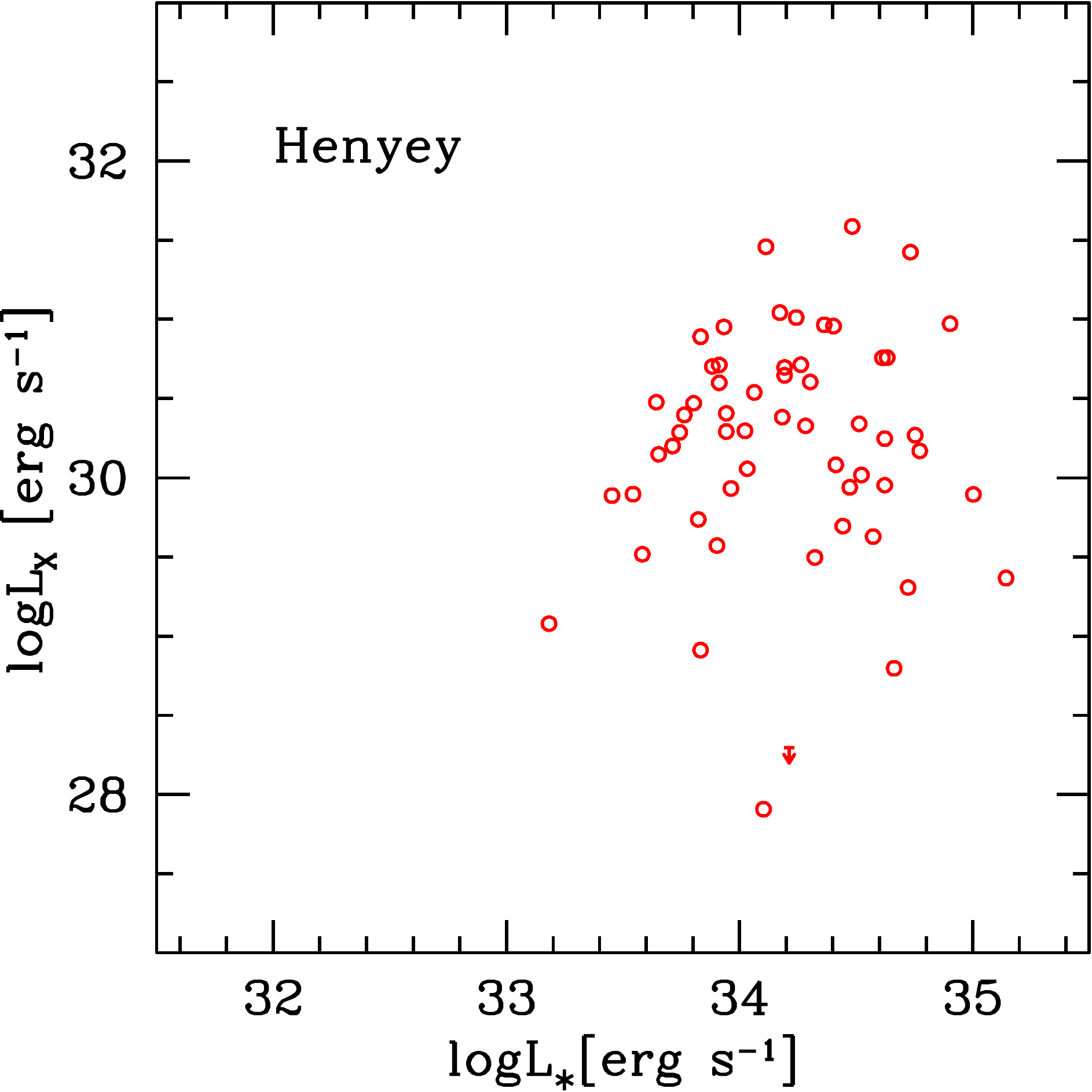}
	\caption{The correlation between $L_X$ and $L_\ast$ for fully convective ({\it top left}, $L_X\propto L_\ast^a$ with $a=0.93\pm0.04$) and for partially convective PMS stars ({\it top right}, $a=0.33\pm0.09$).  Points are coloured as in Figure \ref{HRdiagram}. Downward pointing arrows are stars with a $L_X$ upper limit. The almost linear relationship between $L_X$ and $L_\ast$ is maintained for Hayashi track stars ({\it bottom left}, $a=0.92\pm0.04$), however, $\sim$90\% of the such stars are fully convective.  If we instead consider only Hayashi track stars with radiative cores ({\it bottom middle}) the gradient of the correlation is only $a=0.61\pm0.08$. In all cases the probability of there not being a correlation, from generalised Kendall's $\tau$ tests, is $<$5e-5, with the exception of Henyey track PMS stars ({\it bottom right}) for which there is no correlation. Figures modified from \citet{gre16}.}
	\label{logLX_logLstar}
\end{figure*}

The reason why Henyey track PMS stars have a lower average $\log(L_X/L_\ast)$ compared to Hayashi track PMS stars, is because many of them are under-luminous in X-rays for their bolometric luminosity.  This can be seen from Figure \ref{logLX_logLstar}.  If we consider fully convective stars only there is an almost linear relationship between $L_X$ and $L_\ast$, with $L_X\propto L_\ast^a$, $a=0.93\pm0.04$, and $P(0)<$5e-5.\footnote{Throughout this work linear regression fits are calculated using the expectation-maximisation algorithm, which accounts for the $L_X$ upper limits in the sample, within the ASURV (Astronomy SURVival analysis) package \citep{iso86}. Quoted $P(0)$ values are probabilities that correlations do not exist calculated from generalised Kendall's $\tau$ tests, also carried out with ASURV.} The exponent drops to $a=0.33\pm0.09$ [$P(0)<$5e-5] for partially convective stars.  If we instead consider Hayashi track stars, an almost linear correlation is again found with $a=0.92\pm0.04$ [$P(0)<$5e-5]. However, this hides the fact that the majority of Hayashi track stars are fully convective.  For partially convective Hayashi track stars (the blue points in Figure \ref{logLX_logLstar}) the exponent is smaller: $a=0.61\pm0.08$ [$P(0)<$5e-5].  For Henyey track stars there is no correlation between $L_X$ and $L_\ast$ [$P(0)=$0.52], with many such stars having $L_X$ well below what would be expected given their $L_\ast$.  Once stars have developed substantial radiative interiors, their X-ray emission appears to decay.        

\begin{figure*}
	\centering
	\includegraphics[width=0.33\linewidth]{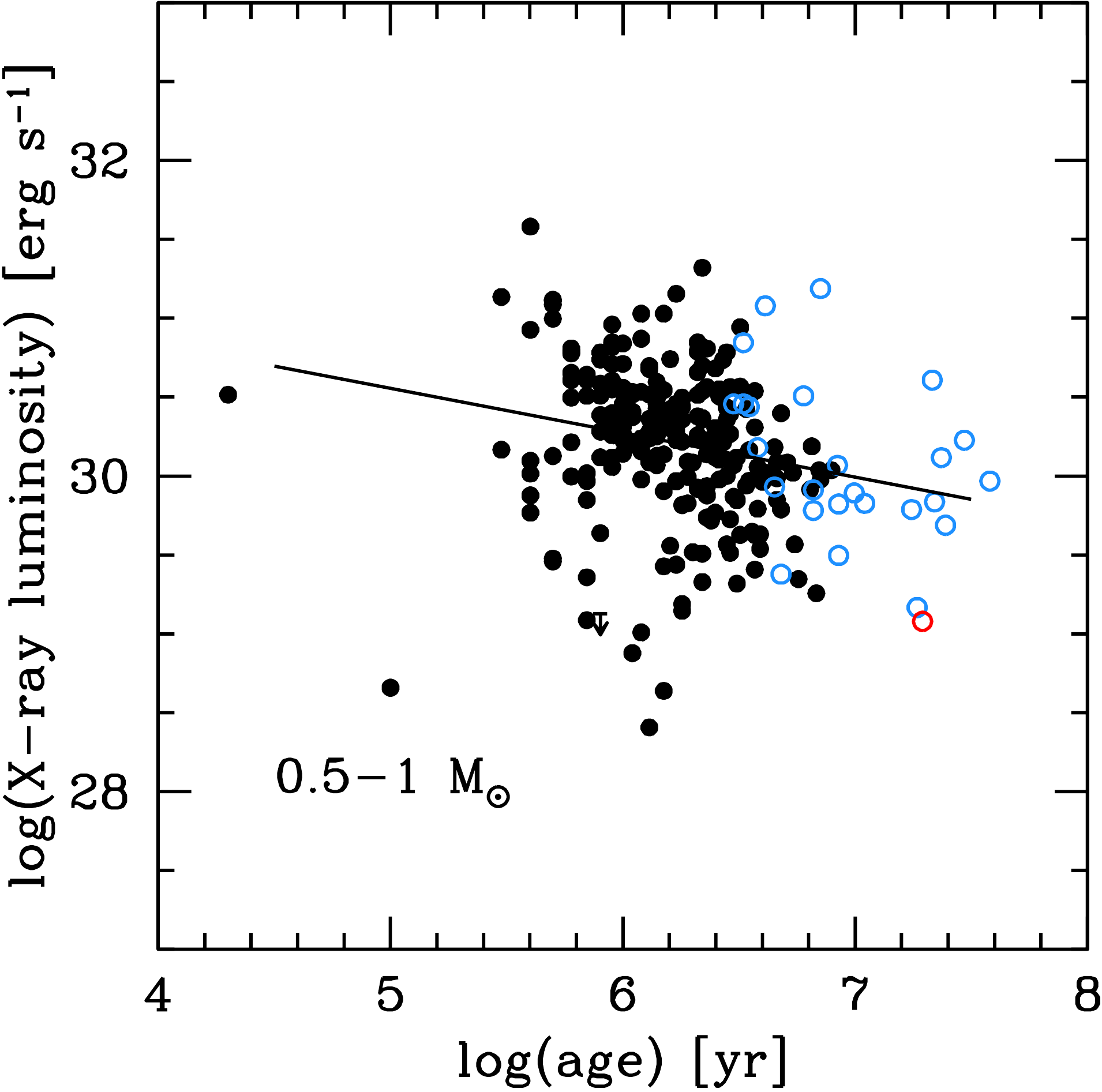}
	\includegraphics[width=0.33\linewidth]{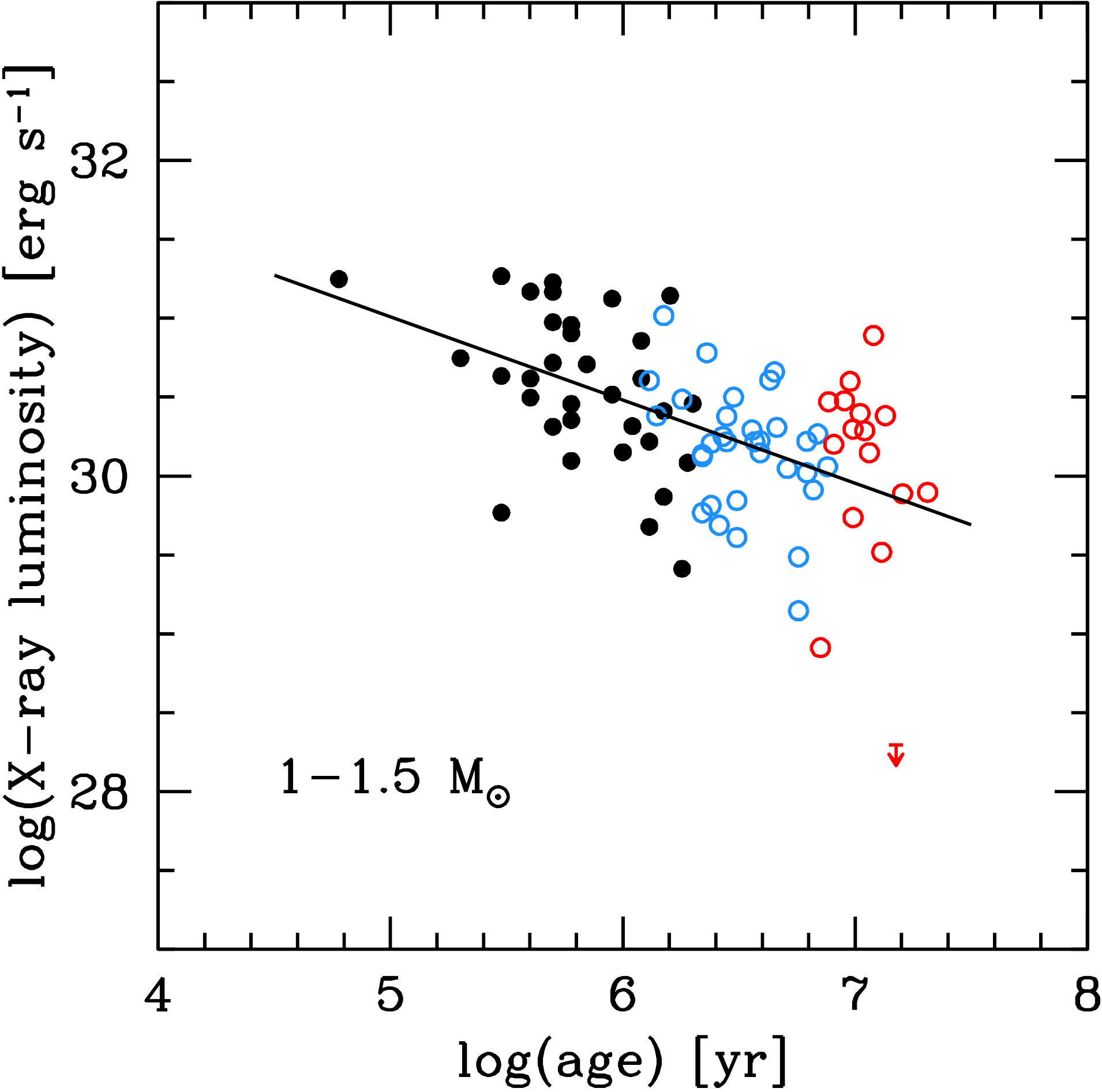}\\
	\includegraphics[width=0.33\linewidth]{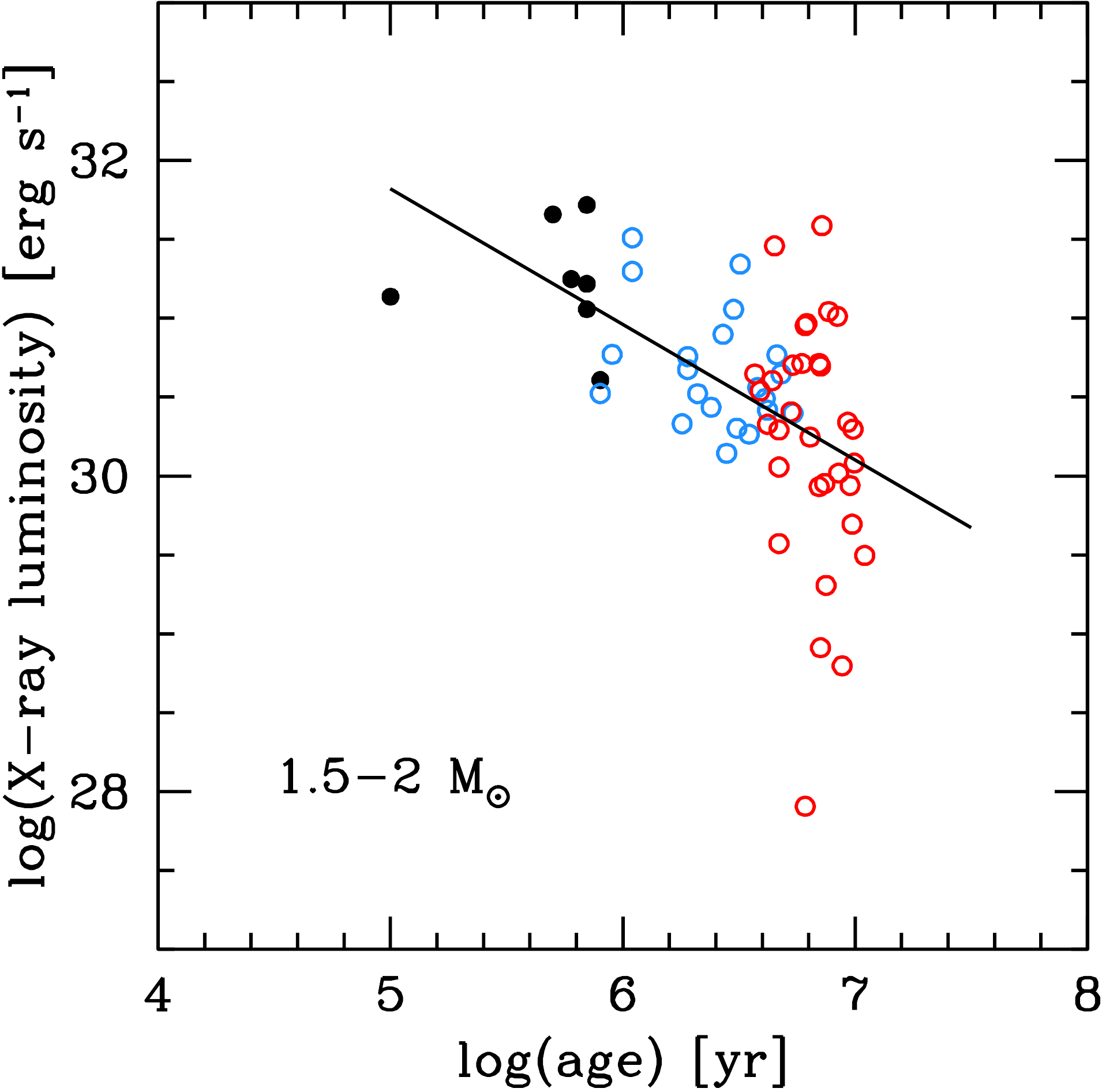}
	\includegraphics[width=0.33\linewidth]{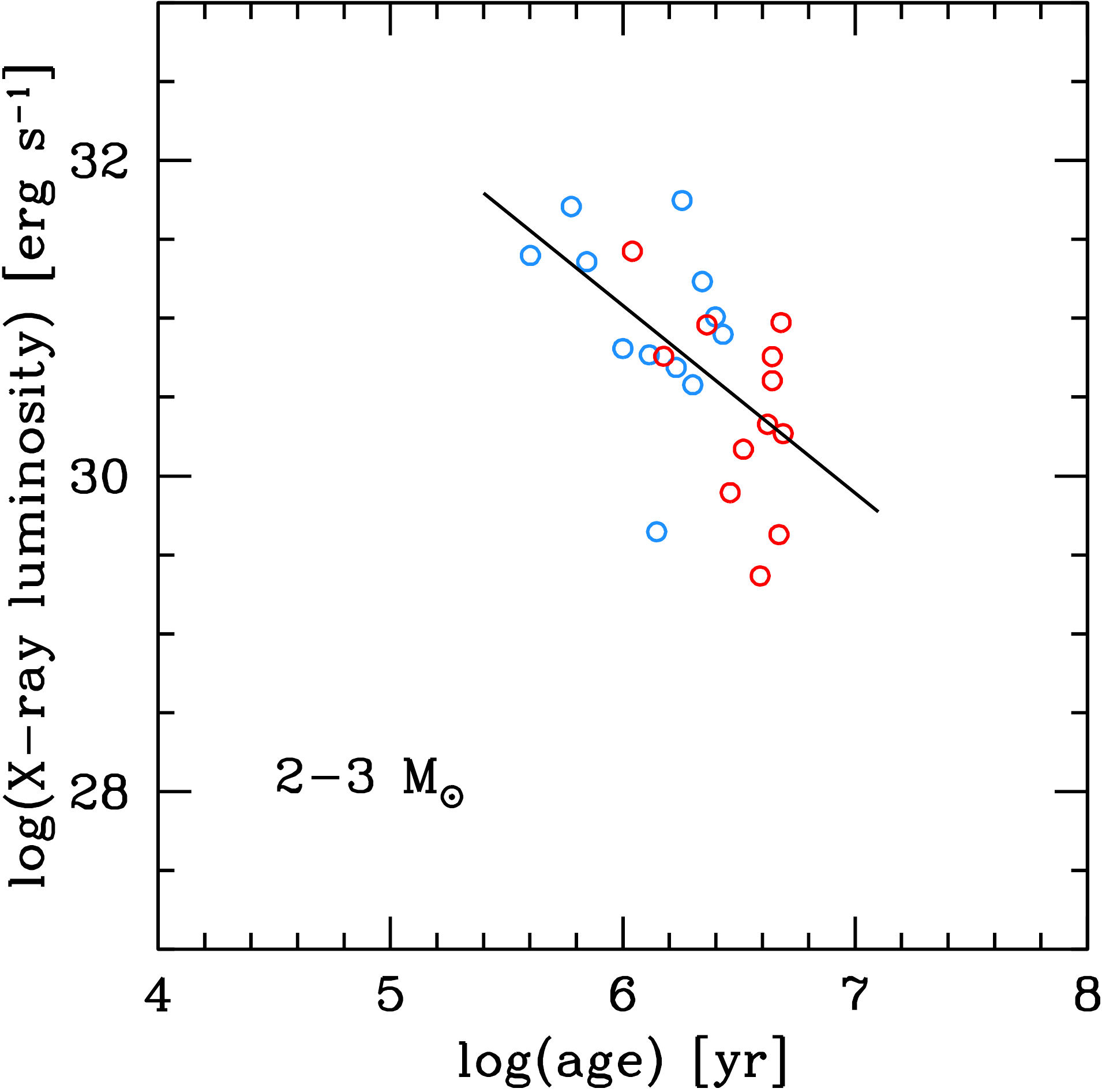}
	\caption{The decay of $L_X$ with age for stars of the indicated mass.  Points are coloured as in Figure \ref{HRdiagram}. Downward pointing arrows are stars with a $L_X$ upper limit. $L_X$ decays faster with age, $t$, for higher mass stars with $L_X\propto t^a$ where $a=-0.28\pm0.08, -0.53\pm0.10, -0.86\pm0.19$, and $-1.19\pm0.35$ for 0.5-1, 1-1.5, 1.5-2, and 2-3$\,{\rm M}_\odot$ respectively. $P(0)=$0.0059 for the 2-3$\,{\rm M}_\odot$ mass range and $<$5e-5 for the others. Figure from \citet{gre16}.}
	\label{logLX_logage}
\end{figure*}

\begin{figure}
	\centering
	\includegraphics[width=0.8\linewidth]{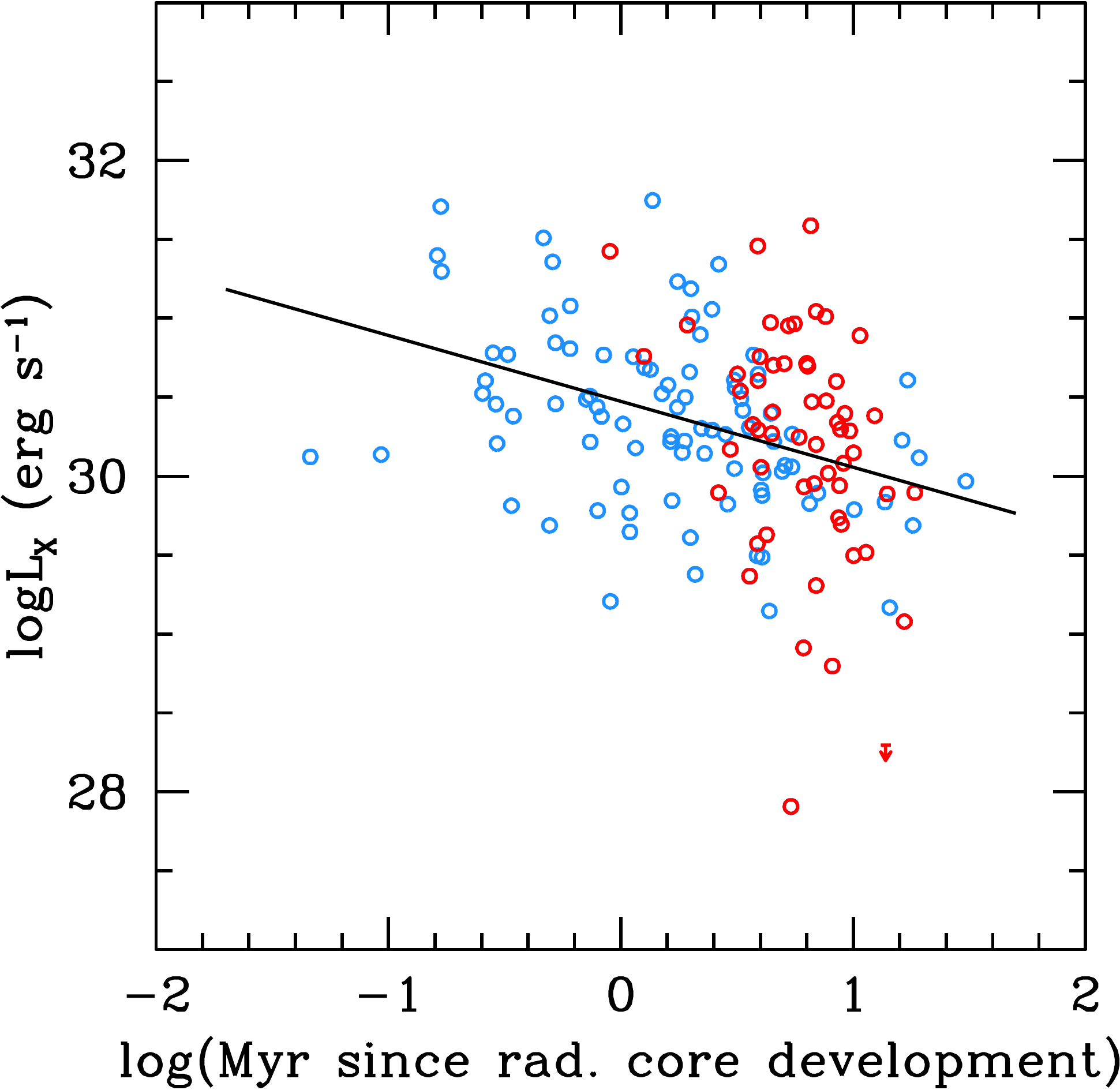}
	\caption{The correlation between $L_X$ and time since radiative core development, $L_X\propto t_{\rm since}^a$ with $a=-0.42\pm0.09$ and $P(0)<$5e-5. Blue/red points are stars on Hayashi/Henyey tracks. The downward pointing arrow is a star with a $L_X$ upper limit. The longer a PMS star has spent with a radiative core the less X-ray luminous it becomes. Figure from \citet{gre16}.}
	\label{logLX_tsince}
\end{figure}

X-ray emission decays with age for all PMS stars \citep{pre05age}.  Considering mass-stratified samples, $L_X$ is found to decay faster with age for higher mass PMS stars, see Figure \ref{logLX_logage}.  In higher mass bins there is a greater proportion of partially convective (red/blue points in Figure \ref{logLX_logage}) and Henyey track (red points) stars compared to fully convective objects (black points). This again suggests that changes in the stellar internal structure, and the evolution of PMS stars from Hayashi to Henyey tracks and the associated development of a substantial radiative core, reduces the coronal X-ray emission. We also find that the longer a star has spent with a radiative core, $t_{\rm since}$, the weaker its X-ray emission becomes with $L_X\propto t_{\rm since}^{-2/5}$, see Figure \ref{logLX_tsince}.  However, we do not find any correlation between $L_X$ and radiative core mass or radius \citep{gre16}.      

In the following section we argue that the decay in X-ray emission with substantial radiative core growth is consistent with the lack of X-ray detections of main sequence A-type stars. 


\section{Do intermediate mass stars lose their coronae on the PMS?}\label{Atype}
The young early K-type to late-G type PMS stars in our sample will evolve into main sequence A-type stars \citep{sie00} which lack outer convective zones.  The decay of their X-ray emission with substantial radiative core growth, discussed in the previous section, is consistent with the lack of X-ray detections of main sequence A-type stars.

\begin{figure}
	\centering
	\includegraphics[width=0.8\linewidth]{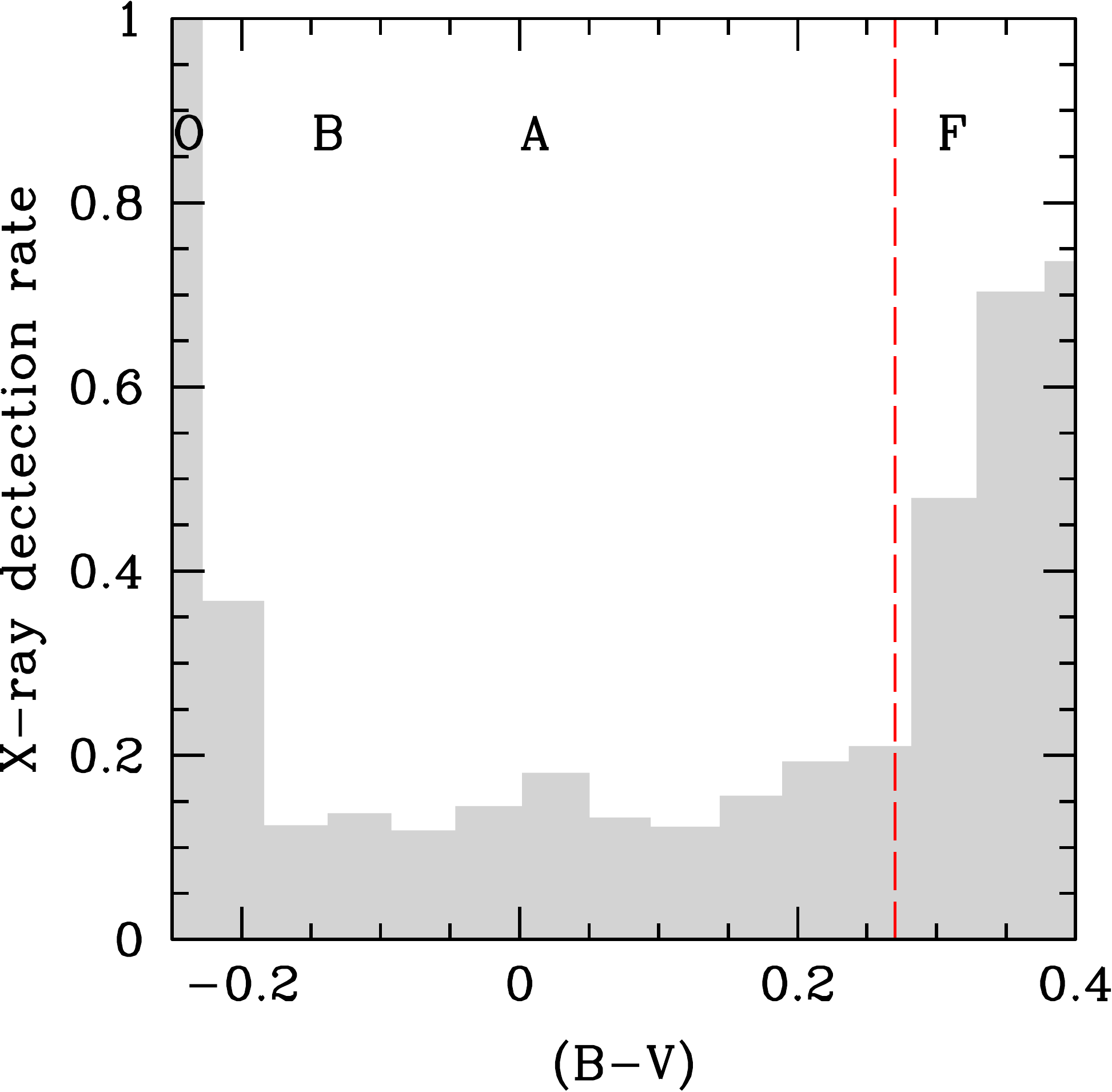}
	\caption{The X-ray detection rate vs the $(B-V)$ colour of bright stars in the ROSAT All-Sky Survey (flux limited sample). Moving right from the dashed red line corresponds to increasing outer convective zone depth. The approximate spectral type boundaries are indicated. Most A-type stars go undetected in X-rays.  Of those that are, almost all are known binaries with later spectral type companions. Figure modified from \citet{sch09}, following \citet{sch07}.}
	\label{rosat}
\end{figure}

Figure \ref{rosat} shows the X-ray detection rate of early type stars, with the approximate spectral type boundaries indicated. Only 10-15\% of A-type stars are detected in X-rays, and of these, almost all have later spectral type binary companions that are the suspected source of the X-ray emission \citep{sch07}.  The X-ray detection rate increases moving through F-type stars and with increasing outer convective zone depth. The X-ray detection rate of F-type stars is almost 100\% if a volume limited sample is considered instead of the flux limited sample plotted in Figure \ref{rosat} \citep{sch04}.  

There are a few (late) A-type stars where weak coronal X-ray emission has been detected (e.g. \citealt{rob10,gun12}).  However, at a level of $\log L_X \approx$ 26.5--28 (with $L_X$ in units of $\,{\rm erg\,s^{-1}}$) this is well below what is typical of their young, progenitor, PMS stars (see Figure \ref{logLX_logLstar}).  Given our discussion in section \ref{decay}, with the highest mass PMS stars showing the steepest decay in $L_X$ with age, and the reduction in $L_X$ with time since radiative core development, it seems we are observing the loss of the coronal X-ray emission from stars that will evolve to X-ray dark (or at least very weak X-ray emitting) main sequence A-type stars.    


\section{Conclusions}\label{conclusions}
We have compared the X-ray emission properties of fully and partially convective PMS stars and of PMS stars on Hayashi and Henyey tracks in the HR diagram. We have found that the growth of the radiative core plays an important role in determining the behaviour of X-ray emission from PMS stars. Our results can be summarised as follows:
\begin{itemize}
\item Partially convective PMS stars have a lower average $\log(L_X/L_\ast)$ compared to fully convective PMS stars.  The deficit is larger for Henyey track PMS stars compared to those on Hayashi tracks in the HR diagram.
\item The lower average $\log(L_X/L_\ast)$ for partially convective PMS stars is driven by a decay in $L_X$ with radiative core growth.  $L_X$ reduces with age and does so faster for higher mass PMS stars.
\item The longer PMS stars have spent with radiative cores the less X-ray luminous they become.
\item The (roughly) linear correlation between $L_X$ and $L_\ast$ is only valid for fully convective PMS stars.  The exponent of the correlation (i.e. the value $a$ where $L_X\propto L_\ast^a$) is less when considering partially convective PMS stars.
\item There is no correlation between $L_X$ and $L_\ast$ for Henyey track PMS stars (which have mostly radiative interiors) due to many having $L_X$ values well below what is typical of Hayashi track PMS stars of similar $L_\ast$.
\item Stars which are the progenitors of (X-ray undetected) main sequence A-type stars are already losing their coronal X-ray emission during their PMS evolution. 
\end{itemize}
More details, a greater discussion of the sample of stars used in our work, and additional correlations can be found in \citet{gre16}.

The reduction in $L_X$ with substantial radiative core development may be linked to the observationally inferred transition in the large-scale magnetic field topologies of PMS stars.  Zeeman-Doppler imaging observations (e.g. \citealt{don11b,don11a,don12}) have revealed that (accreting) PMS stars are born with simple and axisymmetric large-scale magnetic fields, that are well-described by a tilted dipole plus a tilted octupole component \citep{gre11}, which become more dominantly octupolar with age \citep{gre14}. As PMS stars develop substantial radiative cores their large-scale magnetic fields become highly multipolar and non-axisymmetric \citep{gre12}. If the large-scale magnetic field is able to contain X-ray emitting coronal plasma\footnote{The X-ray emitting structures within the coronae of PMS stars are thought to extend to at least 2 stellar radii \citep{arg16}, far in excess of the more compact X-ray emitting regions resolved on the Sun.}, then a transition from mostly low-order multipole / axisymmetric to a mostly high-order multipole / non-axisymmetric magnetic field would correspond to a decrease in the X-ray emitting volume. In turn, this reduces the volume emission measure and therefore the X-ray luminosity.


\section*{Acknowledgements}
{SGG acknowledges support from the Science \& Technology Facilities Council (STFC) via an Ernest Rutherford Fellowship [ST/J003255/1]. CLD acknowledges support from STFC via a PhD studentship and additional funding via the STFC Studentship Enhancement Programme [ST/J500744/1], and support from the ERC Starting Grant "ImagePlanetFormDiscs" (Grant Agreement No. 639889).}

\bibliographystyle{cs19proc}
\bibliography{sgregory_xray_v2.bib}

\begin{thebibliography}{33}
\providecommand{\natexlab}[1]{#1}

\bibitem[\protect\astroncite{{Argiroffi} \emph{et~al.}}{2016}]{arg16}
{Argiroffi}, C., {Caramazza}, M., {Micela}, G., {Sciortino}, S., {Moraux}, E.,
  \emph{et~al.} 2016, \aap, 589, A113.

\bibitem[\protect\astroncite{{Broos} \emph{et~al.}}{2013}]{bro13}
{Broos}, P.~S., {Getman}, K.~V., {Povich}, M.~S., {Feigelson}, E.~D.,
  {Townsley}, L.~K., \emph{et~al.} 2013, \apjs, 209, 32.

\bibitem[\protect\astroncite{{Currie} \emph{et~al.}}{2009}]{cur09}
{Currie}, T., {Evans}, N.~R., {Spitzbart}, B.~D., {Irwin}, J., {Wolk}, S.~J.,
  \emph{et~al.} 2009, \aj, 137, 3210.

\bibitem[\protect\astroncite{{Donati}
  \emph{et~al.}}{2011{\natexlab{a}}}]{don11b}
{Donati}, J.-F., {Bouvier}, J., {Walter}, F.~M., {Gregory}, S.~G., {Skelly},
  M.~B., \emph{et~al.} 2011{\natexlab{a}}, \mnras, 412, 2454.

\bibitem[\protect\astroncite{{Donati}
  \emph{et~al.}}{2011{\natexlab{b}}}]{don11a}
{Donati}, J.-F., {Gregory}, S.~G., {Alencar}, S.~H.~P., {Bouvier}, J.,
  {Hussain}, G., \emph{et~al.} 2011{\natexlab{b}}, \mnras, 417, 472.

\bibitem[\protect\astroncite{{Donati} \emph{et~al.}}{2012}]{don12}
{Donati}, J.-F., {Gregory}, S.~G., {Alencar}, S.~H.~P., {Hussain}, G.,
  {Bouvier}, J., \emph{et~al.} 2012, \mnras, 425, 2948.

\bibitem[\protect\astroncite{{Feigelson} \emph{et~al.}}{2003}]{fei03}
{Feigelson}, E.~D., {Gaffney}, J.~A., III, {Garmire}, G., {Hillenbrand}, L.~A.,
  \& {Townsley}, L. 2003, \apj, 584, 911.

\bibitem[\protect\astroncite{{Feigelson} \emph{et~al.}}{2013}]{fei13}
{Feigelson}, E.~D., {Townsley}, L.~K., {Broos}, P.~S., {Busk}, H.~A., {Getman},
  K.~V., \emph{et~al.} 2013, \apjs, 209, 26.

\bibitem[\protect\astroncite{{Flaccomio} \emph{et~al.}}{2003}]{fla03}
{Flaccomio}, E., {Damiani}, F., {Micela}, G., {Sciortino}, S., {Harnden},
  F.~R., Jr., \emph{et~al.} 2003, \apj, 582, 398.

\bibitem[\protect\astroncite{{Folsom} \emph{et~al.}}{2016}]{fol16}
{Folsom}, C.~P., {Petit}, P., {Bouvier}, J., {L{\`e}bre}, A., {Amard}, L.,
  \emph{et~al.} 2016, \mnras, 457, 580.

\bibitem[\protect\astroncite{{Getman} \emph{et~al.}}{2010}]{get10}
{Getman}, K.~V., {Feigelson}, E.~D., {Broos}, P.~S., {Townsley}, L.~K., \&
  {Garmire}, G.~P. 2010, \apj, 708, 1760.

\bibitem[\protect\astroncite{{Gregory} \emph{et~al.}}{2016}]{gre16}
{Gregory}, S.~G., {Adams}, F.~C., \& {Davies}, C.~L. 2016, \mnras, 457, 3836.

\bibitem[\protect\astroncite{{Gregory} \& {Donati}}{2011}]{gre11}
{Gregory}, S.~G. \& {Donati}, J.-F. 2011, Astronomische Nachrichten, 332, 1027.

\bibitem[\protect\astroncite{{Gregory} \emph{et~al.}}{2012}]{gre12}
{Gregory}, S.~G., {Donati}, J.-F., {Morin}, J., {Hussain}, G.~A.~J., {Mayne},
  N.~J., \emph{et~al.} 2012, \apj, 755, 97.

\bibitem[\protect\astroncite{{Gregory} \emph{et~al.}}{2014}]{gre14}
{Gregory}, S.~G., {Donati}, J.-F., {Morin}, J., {Hussain}, G.~A.~J., {Mayne},
  N.~J., \emph{et~al.} 2014, In \emph{Magnetic Fields throughout Stellar
  Evolution}, edited by P.~{Petit}, M.~{Jardine}, \& H.~C. {Spruit}, \emph{IAU
  Symposium}, vol. 302, pp. 40--43.

\bibitem[\protect\astroncite{{G{\"u}nther} \emph{et~al.}}{2012}]{gun12}
{G{\"u}nther}, H.~M., {Wolk}, S.~J., {Drake}, J.~J., {Lisse}, C.~M., {Robrade},
  J., \emph{et~al.} 2012, \apj, 750, 78.

\bibitem[\protect\astroncite{{Herczeg} \& {Hillenbrand}}{2014}]{her14}
{Herczeg}, G.~J. \& {Hillenbrand}, L.~A. 2014, \apj, 786, 97.

\bibitem[\protect\astroncite{{Hillenbrand} \emph{et~al.}}{2013}]{hil13}
{Hillenbrand}, L.~A., {Hoffer}, A.~S., \& {Herczeg}, G.~J. 2013, \aj, 146, 85.

\bibitem[\protect\astroncite{{Isobe} \emph{et~al.}}{1986}]{iso86}
{Isobe}, T., {Feigelson}, E.~D., \& {Nelson}, P.~I. 1986, \apj, 306, 490.

\bibitem[\protect\astroncite{{Mayne}}{2010}]{may10}
{Mayne}, N.~J. 2010, \mnras, 408, 1409.

\bibitem[\protect\astroncite{{McGill} \emph{et~al.}}{1978}]{mcg78}
{McGill}, R., {Tukey}, J.~W., \& {Larsen}, W.~A. 1978, The American
  Statistician, 32, 12.

\bibitem[\protect\astroncite{{Pecaut} \& {Mamajek}}{2013}]{pec13}
{Pecaut}, M.~J. \& {Mamajek}, E.~E. 2013, \apjs, 208, 9.

\bibitem[\protect\astroncite{{Peres} \emph{et~al.}}{2000}]{per00}
{Peres}, G., {Orlando}, S., {Reale}, F., {Rosner}, R., \& {Hudson}, H. 2000,
  \apj, 528, 537.

\bibitem[\protect\astroncite{{Preibisch} \& {Feigelson}}{2005}]{pre05age}
{Preibisch}, T. \& {Feigelson}, E.~D. 2005, \apjs, 160, 390.

\bibitem[\protect\astroncite{{Preibisch} \emph{et~al.}}{2005}]{pre05}
{Preibisch}, T., {Kim}, Y.-C., {Favata}, F., {Feigelson}, E.~D., {Flaccomio},
  E., \emph{et~al.} 2005, \apjs, 160, 401.

\bibitem[\protect\astroncite{{Rebull} \emph{et~al.}}{2006}]{reb06}
{Rebull}, L.~M., {Stauffer}, J.~R., {Ramirez}, S.~V., {Flaccomio}, E.,
  {Sciortino}, S., \emph{et~al.} 2006, \aj, 131, 2934.

\bibitem[\protect\astroncite{{Robrade} \& {Schmitt}}{2010}]{rob10}
{Robrade}, J. \& {Schmitt}, J.~H.~M.~M. 2010, \aap, 516, A38.

\bibitem[\protect\astroncite{{Schmitt}}{2009}]{sch09}
{Schmitt}, J. 2009, In \emph{High Resolution X-ray Spectroscopy: Towards IXO},
  p. E38.

\bibitem[\protect\astroncite{{Schmitt} \& {Liefke}}{2004}]{sch04}
{Schmitt}, J.~H.~M.~M. \& {Liefke}, C. 2004, \aap, 417, 651.

\bibitem[\protect\astroncite{{Schr{\"o}der} \& {Schmitt}}{2007}]{sch07}
{Schr{\"o}der}, C. \& {Schmitt}, J.~H.~M.~M. 2007, \aap, 475, 677.

\bibitem[\protect\astroncite{{Siess} \emph{et~al.}}{2000}]{sie00}
{Siess}, L., {Dufour}, E., \& {Forestini}, M. 2000, \aap, 358, 593.

\bibitem[\protect\astroncite{{Stelzer} \emph{et~al.}}{2012}]{ste12}
{Stelzer}, B., {Preibisch}, T., {Alexander}, F., {Mucciarelli}, P.,
  {Flaccomio}, E., \emph{et~al.} 2012, \aap, 537, A135.

\bibitem[\protect\astroncite{{Wright} \emph{et~al.}}{2011}]{wri11}
{Wright}, N.~J., {Drake}, J.~J., {Mamajek}, E.~E., \& {Henry}, G.~W. 2011,
  \apj, 743, 48.

\end{thebibliography}

\end{document}